\documentclass[final,5p,times]{elsarticle}

\usepackage{graphicx}

\usepackage{amssymb}
\usepackage{amsmath}
\usepackage{color}

\journal{Synthetic Metals}

\begin{document}

\begin{frontmatter}



\title{Discrete differential geometry and the properties of conformal two-dimensional materials}

\author[label1]{Salvador Barraza-Lopez}
\ead[e-mail:]{sbarraza@uark.edu}

\address[label1]{Department of Physics. University of Arkansas. Fayetteville AR, 72701. USA}

\begin{abstract}
Two-dimensional materials were first isolated no longer than ten years ago, and a comprehensive understanding of their properties under non-planar shapes is still being developed. Strictly speaking, the theoretical study of the properties of graphene and other two-dimensional materials is the most complete for planar structures and for structures with small deformations from planarity. The opposite limit of large deformations is yet to be studied comprehensively but that limit is extremely relevant because it determines material properties near the point of failure. We are exploring uses for discrete differential geometry within the context of graphene and other two-dimensional materials, and these concepts appear promising in linking materials properties to shape regardless of how large a given material deformation is. A brief account of additional contributions arising from our group to two-dimensional materials that include graphene, stanene and phosphorene is provided towards the end of this manuscript.
\end{abstract}

\end{frontmatter}


Geometry is a pillar of Science, and many physical theories are decidedly geometric \cite{Nelson}. This manuscript provides an overview of recent developments  towards linking the properties of two-dimensional materials to a given arbitrary shape, where {\em shape} is understood as the local two-dimensional geometry of atom-thin materials that are embedded on a three-dimensional space. The unifying point of the description concerns the introduction of a discrete geometry to deal with two-dimensional materials while fully preserving their atomistic information.

Thus, we showcase here a set of geometrical principles that apply to nets, where a {\em net} is a discrete surface or a mesh. We identify two-dimensional materials with {\em meshes}, and apply precepts from a branch of Mathematics \cite{Math1} that deals with discrete surfaces. We have presented a number of results in linking this geometry to materials properties already \cite{usPRB,usSSC,ACSNano,usRapid,newest}. This subject has contributions from other teams as well \cite{Monica,TomanekPRB,Kamien}.

I acknowledge Drs. Pacheco SanJuan, Wang, Harriss, Rivero, Vanevi\'c, and Terrones' contributions to this ongoing work. I also thank students Sloan, Horvath, Utt, Pour-Imani, Mehboudi, and Klee for their contributions at different stages. I am grateful to many colleagues for their observations and encouragement; most particularly to Mar{\'\i}a A. H. Vozmediano.

Graphene and other 2-D materials provide a stage to further our understanding of Physics. Perhaps the most natural connection to be studied concerns the creation of gauge fields on effective Dirac particles in 2+1 dimensions as the geometry evolves from a reference, planar shape \cite{Pereira1,GuineaNatPhys2010,vozmediano}, to be addressed next.

The starting point for us was the analysis of strain created by a scanning tunneling microscope (STM) on graphene \cite{usold}. There is a theory laid out on a structural continuum \cite{Ando2002,Pereira1,GuineaNatPhys2010,vozmediano,Wakker,Kim,CarrilloBastos,Sandler} that correlates structural deformations to mechanically-induced gauges on Dirac fermions in 2+1 dimensions. These effective Dirac fermions arise from a first nearest neighbor tight-binding description of $\pi-$electrons on graphene at low energies. Changes in distances arising from a structural deformation are estimated from a continuum model of the distortion, and these changes in distances alter the magnitude of the tight-binding hopping terms locally.

The formulation is inherently semi-classical, in the sense that the underlying dynamics is that of pseudospins (which strictly speaking are only valid on the ideal non-deformed crystalline structure) and the gauge fields produced by mechanical deformations induce local modifications to the $\pi$-electron pseudospin Hamiltonian. We estimated gauge fields employing that formalism \cite{Pereira1,GuineaNatPhys2010,vozmediano} but this question quickly came up:
\begin{enumerate}
\item{}An STM can tell individual atoms. Can one rewrite the theory expressed on a continuum structure to reflect such atomistic nature? What do we learn when the theory is laid out this way that is different from the continuum formalism?
\end{enumerate}

This paper contains three sections that are somehow independent: (1) Its main thrust is the description of the coupling of finite displacements to a semiclassical pseudospin dynamics of  $\pi$-electrons on graphene in which we attempt to provide answers to the two questions above (pages 2-5). (2) We then provide a description of a discrete geometry that applies to arbitrary two-dimensional materials (pages 5-7). (3) The document ends by briefly mentioning other developments in graphene and other materials in which we have been involved (pages 8-10).

\section{A lattice gauge field theory for Dirac fermions in graphene}

The interplay among the electronic  and mechanical properties of graphene membranes remains under experimental and theoretical investigation \cite{Pereira1,GuineaNatPhys2010,vozmediano,Nature2007,McEuen1,Kitt2012,Ando2002,deJuanPRL2012}, and an insightful picture of the effects of deformations employs gauge fields that influence the dynamics of charge carriers \cite{Pereira1,GuineaNatPhys2010,vozmediano,Kitt2012,Ando2002}.

The formulation is inherently semi-classical and takes pseudospin hamiltonians as the main object, which strictly speaking are only valid on the ideal non-deformed crystalline structure, with gauge fields arising from slow-varying mechanical deformations providing local modifications to the said Hamiltonian.

But graphene can sustain elastic deformations as large as 20\% \cite{nature457_706} and using this picture, the resulting pseudo-magnetic fields are much larger than those magnetic fields available in state-of-the-art experimental facilities. The presence of a  pseudo-magnetic field is observed via broad Landau levels (LLs) in strained graphene nanobubbles on a metal substrate \cite{Crommie}. In addition to the pseudo-magnetic vector potential $\mathbf{A}_s$, strain also induces a scalar deformation potential $E_s$ \cite{Ando2002,deJuanPRB,YWSon} that affects the electron dynamics in non-trivial ways. Part of our motivation was to reconcile the experimental results that can be obtained when the lattice is largely deformed, with a theory that by construction applies to small deformations. What we accomplished is a close view at the inner workings of this theory that has led to unique insights, and a quantitative understanding of ``slowly varying deformations'' within the context of this theory. Our formulation brings to the spotlight some of the inherent assumptions on the prevailing theoretical framework.

The underlying assumptions of the theory expressed on a structural continuum are expressed in the following sentence: ``If a mechanical strain varies smoothly on the scale of interatomic distances, it does not break sublattice symmetry but rather deforms the Brillouin zone in such a way that the Dirac cones located in graphene at points $K$ and $K'$ are shifted in opposite directions ~\cite{GuineaNatPhys2010}.''

Previous statement tells us that --{\em provided strain preserves sublattice symmetry}-- one can understand the effects of mechanical strain on the electronic structure in terms of a semiclassical approach, in which mechanical strain induces the spatially-varying gauge fields $B_s(\mathbf{r})=\nabla \times A_s(\mathbf{r})$ and $E_s(\mathbf{r})$ into a spatially-varying pseudospin Hamiltonian $\mathcal{H}_{ps}(\mathbf{q},\mathbf{r})$, where $\mathcal{H}_{ps}(\mathbf{q})$ is the low-energy expansion of the Hamiltonian in reciprocal space in the absence of strain. This semiclassical approximation is justified when the strain extends over many unit cells and it preserves sublattice symmetry \cite{GuineaNatPhys2010,vozmediano,Ando2002}, and many of the equations on this Section will help us keep track of said sublattice symmetry.

Evidently, it is possible to determine the electronic properties directly from a tight-binding Hamiltonian $\mathcal{H}$ in real space, without resorting to the semiclassical approximation and without imposing a sublattice symmetry {\em a priori}. That is, while the semiclassical $\mathcal{H}_{ps}(\mathbf{q},\mathbf{r})$ is defined in reciprocal space (thus assuming some reasonable preservation of crystalline order), the tight-binding Hamiltonian $\mathcal{H}$ in real space is more general and can be used for membranes with arbitrary spatial distribution and magnitude of the strain.

In the previous formulation of the theory both $\mathbf{A}_s$ and $E_s$ are expressed in terms of a {\em continuous} displacement field $\mathbf{u}(x,y)$ obtained within first-order continuum elasticity (CE) \cite{Pereira1,GuineaNatPhys2010,vozmediano,Ando2002}. It is not possible to assess sublattice symmetry on a continuum media, and therefore proper phase conjugation of pseudospin Hamiltonians becomes an implicit assumption of that theory.
\begin{figure}[tb]
\includegraphics[width=0.45\textwidth]{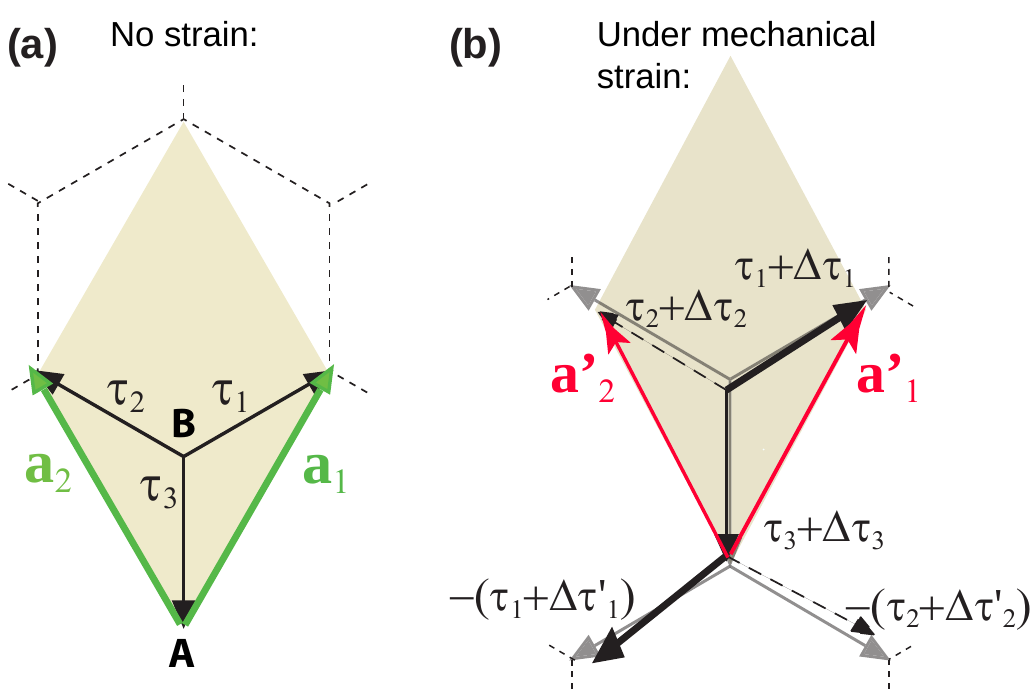}
\caption{(a) Definitions of geometrical parameters in a unit cell. (b) Sublattice symmetry relates to how {\em pairs} of nearest-neighbor vectors (either in thick, or dashed lines) are modified due to strain. These vectors change by $\Delta \mathbf{\tau}_j$ and $\Delta \mathbf{\tau}_j'$ upon strain ($j=1,2$). Relative displacements of neighboring atoms lead to modified lattice vectors; the choice of renormalized lattice vectors will be unique {\em only} to the extent to which sublattice symmetry is preserved: $\Delta \mathbf{\tau}_j'\simeq \Delta \mathbf{\tau}_j$.}\label{fig:F2}
\end{figure}

But the only way to know whether the strain preserves sublattice symmetry \cite{GuineaNatPhys2010} implies analyzing relative atomic displacements  in arbitrary structural distortions that could be captured directly from experiment \cite{MolecularGraphene}, or from molecular dynamics simulations.

Let us start by considering the unit cell before (Fig.~\ref{fig:F2}(a)) and  after arbitrary strain has been applied (Fig.~\ref{fig:F2}(b)). The lattice vectors and the vectors joining atoms are given by (Fig.~\ref{fig:F2}(a)):
 \begin{equation}\label{eq:defa}
 \mathbf{a}_1=\left(1/2,\sqrt{3}/2\right)a_0,\text{ }\mathbf{a}_2=\left(-{1}/{2},{\sqrt{3}}/{2}\right)a_0,
 \end{equation}
 \begin{equation}\label{eq:deft}
 \boldsymbol{\tau}_1=\left(\frac{\sqrt{3}}{2},\frac{1}{2}\right)\frac{a_0}{\sqrt{3}},\text{ } \boldsymbol{\tau}_2=\left(-\frac{\sqrt{3}}{2},\frac{1}{2}\right)\frac{a_0}{\sqrt{3}},\text{ }
 \boldsymbol{\tau}_3=\left(0,-1\right)\frac{a_0}{\sqrt{3}},
 \end{equation}
 before the deformation takes place. When a deformation is applied (Fig.~\ref{fig:F2}(b)) the two off-diagonal terms making up the pseudospin tight-binding Hamiltonian will be: $$-\sum_{j=1}^3(t+\delta t_j(\Delta \boldsymbol{\tau}_j))e^{i(\boldsymbol{\tau}_j+\Delta\boldsymbol{\tau}_j)\cdot\mathbf{k}},$$ and
 $$-\sum_{j=1}^3(t+\delta t_j(\Delta \boldsymbol{\tau}_j'))e^{i(\boldsymbol{\tau}_j+\Delta\boldsymbol{\tau}_j')\cdot\mathbf{k}},$$
 where $t$ is the hopping term, $\delta t$ is its change upon strain to be explicitly defined later on, and $\mathbf{k}$ is the crystal momentum.

Each local pseudospin Hamiltonian will only have physical meaning when it is properly conjugated, which implies sublattice symmetry holds. This happens at unit cells where:
\begin{equation}\label{eq:applicabilitycondition}
\Delta \boldsymbol{\tau}_j'\simeq\Delta \boldsymbol{\tau}_j \text{ (j=1,2)}.
\end{equation}

One also notes that on this first nearest neighbor picture the diagonal terms are always real even when $\Delta \boldsymbol{\tau}_j'\ne\Delta \boldsymbol{\tau}_j$ (more on this later).

 Condition (\ref{eq:applicabilitycondition}) can be re-expressed in terms of changes of angles $\Delta \alpha_j$ or lengths $\Delta L_j$ for pairs of nearest-neighbor vectors $\boldsymbol{\tau}_j$ and $\boldsymbol{\tau}_j'$
 [$j=1$ is shown in thick solid and $j=2$ in thin dashed lines in Fig.~\ref{fig:F2}(b)]:
\begin{equation}\label{eq:beta}
\small(\boldsymbol{\tau}_j+\Delta  \boldsymbol{\tau}_j)\cdot(\boldsymbol{\tau}_j+\Delta\boldsymbol{\tau}'_j)=
|\boldsymbol{\tau}_j+\Delta\boldsymbol{\tau}_j||\boldsymbol{\tau}_j+\Delta\boldsymbol{\tau}'_j|\cos(\Delta\alpha_j),
\end{equation}
\begin{equation}\label{eq:sign}
\small\text{sgn}(\Delta \alpha_j)=\text{sgn}\left([(\boldsymbol{\tau}_j+\Delta\boldsymbol{\tau}_j)
\times(\boldsymbol{\tau}_j+\Delta\boldsymbol{\tau}'_j)]\cdot \hat{k}\right),\end{equation}
where $\hat{k}$ is a unit vector along the z-axis, $sgn$ is the sign function ($sgn(a)=+1$ if $a\ge 0$ and $sgn(a)=-1$ if $a <0$), and:
\begin{equation}\label{eq:L}
\small
\Delta L_j\equiv |\boldsymbol{\tau}_j+\Delta\boldsymbol{\tau}_j|-|\boldsymbol{\tau}_j+\Delta\boldsymbol{\tau}'_j|.
\end{equation}

Previous expressions indicate that the sublattice symmetry \cite{GuineaNatPhys2010} does not hold {\em a priori}. In the continuum approach, both $\Delta  \boldsymbol{\tau}_j$ and $\Delta  \boldsymbol{\tau}_j'$ are captured at the same point in space using an identical value of the deformation field $\mathbf{u}(x,y)$ and hence the structural aspect just uncovered is hidden. Forcing this symmetry to hold in the lattice depicted at Figure 1b amounts to imposing an artificial mechanical constraint~\cite{Ericksen}, and we re-derive the theory without using that continuum deformation field. The details of the discrete model we developed follow.

\begin{figure}[tb]
\includegraphics[width=0.45\textwidth]{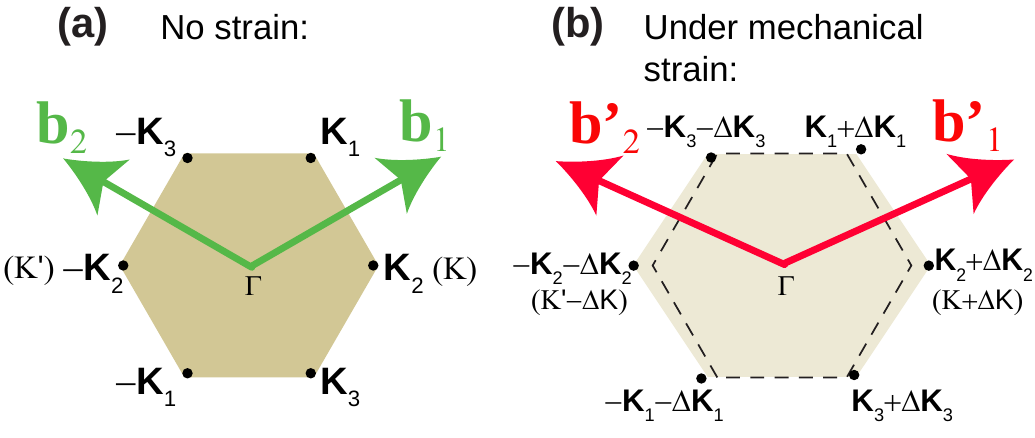}
\caption{First Brillouin zone (a) before and (b) after mechanical strain is applied. The reciprocal lattice vectors are shown,
 as well as the changes of  the high-symmetry points at the corners of the Brillouin zone. Note that independent $K$ points ($K$ and $K'$) move in the opposite directions. The dashed hexagon in (b) represents the boundary of the first Brillouin zone in the absence of strain.}\label{fig:F3}
\end{figure}

In the absence of mechanical strain, the reciprocal lattice vectors $\mathbf{b}_1$ and $\mathbf{b}_2$ are  related to the lattice vectors by \cite{MartinBook}:
\begin{equation}\label{eq:realreciprocal}
\mathcal{B}^T=2\pi\mathcal{A}^{-1},
\end{equation}
where
$\mathcal{A}=\left(\begin{matrix}a_{11}& a_{12}\\a_{21}& a_{22}\end{matrix}\right)$ and $\mathcal{B}=\left(\begin{matrix}b_{11}& b_{12}\\b_{21}& b_{22}\end{matrix}\right)$. We get, with the choice we made for $\mathbf{a}_1$ and $\mathbf{a}_2$:
\begin{equation}
\mathbf{b}_1=\left(1,\frac{1}{\sqrt{3}}\right)\frac{2\pi}{a_0} \text{, and }
\mathbf{b}_2=\left(-1,\frac{1}{\sqrt{3}}\right)\frac{2\pi}{a_0}.
\end{equation}
 The $K-$points on the first Brillouin zone are given by (c.f., Fig.~\ref{fig:F3}(a)):
\begin{equation}
\mathbf{K}_1=\frac{2\mathbf{b}_1+\mathbf{b}_2}{3}, \text{ }\mathbf{K}_2=\frac{\mathbf{b}_1-\mathbf{b}_2}{3} \text{, and } \mathbf{K}_3=-\frac{\mathbf{b}_1+2\mathbf{b}_2}{3},
\end{equation}
and:
\begin{equation}
\mathbf{K}_4=-\mathbf{K}_1,\text{ } \mathbf{K}_5=-\mathbf{K}_2, \text{ and }\mathbf{K}_6=-\mathbf{K}_3.
\end{equation}

The relative positions between atoms change when strain is applied: $\boldsymbol{\tau}_j\to \boldsymbol{\tau}_j+\Delta\boldsymbol{\tau}_j$ ($j=1,2,3)$, and $-\boldsymbol{\tau}_j\to -\boldsymbol{\tau}_j-\Delta\boldsymbol{\tau}_j'$ ($j=1,2$).

How do reciprocal lattice vectors change under mechanical load (to first order)?
Taking Eqn.~\ref{eq:applicabilitycondition} as the starting point, $\Delta \alpha_j$ and $\Delta L_j$ must all be close to zero. In that case we set $\Delta \boldsymbol{\tau}_j'\to \Delta \boldsymbol{\tau}_j$ for j=1,2 and continue our program. We then define:
\begin{equation}
\Delta \mathbf{a}_1\equiv\Delta \boldsymbol{\tau}_1-\Delta \boldsymbol{\tau}_3 \text{, and }
\Delta \mathbf{a}_2\equiv\Delta \boldsymbol{\tau}_2-\Delta \boldsymbol{\tau}_3,
\end{equation}
or in terms of (two-dimensional) components:
\begin{equation}
\Delta \mathcal{A}\equiv
\left(
\begin{matrix}
\Delta \tau_{1x}-\Delta \tau_{3x}& \Delta \tau_{2x}-\Delta \tau_{3x}\\
\Delta \tau_{1y}-\Delta \tau_{3y}& \Delta \tau_{2y}-\Delta \tau_{3y}
\end{matrix}
\right).
\end{equation}
The matrix $\mathcal{A}$ changes to $\mathcal{A}'=\mathcal{A}+\Delta\mathcal{A}$, and we must modify $\mathcal{B}$ so that Eqn.~\eqref{eq:realreciprocal} still holds under mechanical load. To first order in displacements $\mathcal{A}'^{-1}$ becomes:
\begin{equation}\label{eq:correction}
\mathcal{A}'^{-1}=(\mathcal{A}+\Delta\mathcal{A})^{-1}\simeq \mathcal{A}^{-1}-\mathcal{A}^{-1}\Delta\mathcal{A}\mathcal{A}^{-1}.
\end{equation}
By comparing Eqns. (7) and ~\eqref{eq:correction}, the reciprocal lattice vectors in Fig.~\ref{fig:F3}(b) must then be renormalized by:
\begin{equation}
\Delta\mathcal{B}=-2\pi\left(\mathcal{A}^{-1}\Delta\mathcal{A}\mathcal{A}^{-1}\right)^T.
\end{equation}
 This additional term is evident when working directly on the atomic lattice but it was missed in Ref.~\cite{Kitt2012}. Let us now calculate shifts of the $K-$points due to strain. For example, $\mathbf{K}_2$ ($=K$ in Fig.~\ref{fig:F3}(a)) is shifted by:
$$
\Delta K=\Delta\mathbf{K}_2=-\frac{4\pi}{3a_0^2}
\left(\Delta\tau_{1x}-\Delta\tau_{2x},\frac{\Delta \tau_{1x}+\Delta \tau_{2x}-2\Delta \tau_{3x}}{\sqrt{3}}\right),
$$
and using Eqn. (10) one gets $\Delta K'=-\Delta\mathbf{K}_2$, so that the $K$ ($\mathbf{K}_2$) and $K'$ ($-\mathbf{K}_2$) points shift in opposite directions, as expected \cite{GuineaNatPhys2010,castroRMP}.
\subsection{Gauge fields induced by strain}

Equation \eqref{eq:applicabilitycondition} tells whether mechanical strain varies smoothly over interatomic distances  \cite{GuineaNatPhys2010}. This observation provides the rationale for expressing the gauge fields without ever leaving the atomic lattice: When $\Delta \boldsymbol{\tau}_j'\simeq\Delta \boldsymbol{\tau}_j$ at each unit cell a mechanical distortion can be considered ``long-range,'' and the first-order theory is valid. Local gauge fields can be computed as low energy approximations to the following $2\times 2$ pseudospin Hamiltonian:
\begin{equation}\label{eq:tbh}
\left(
\begin{matrix}
E_{s,A} & g^*\\
g & E_{s,B}
\end{matrix}
\right),
\end{equation}
with $g\equiv -\sum_{j=1}^3(t+\delta t_j)e^{i(\boldsymbol{\tau}_j+\Delta\boldsymbol{\tau}_j)\cdot(\mathbf{K}_n+\Delta\mathbf{K}_n+\mathbf{q})}$, and $n=1,...,6$. Keeping exponents to first order we have:
$$
\small
(\boldsymbol{\tau}_j+\Delta\boldsymbol{\tau}_j)\cdot(\mathbf{K}_n+\Delta\mathbf{K}_n+\mathbf{q})\simeq
\boldsymbol{\tau}_j\cdot\mathbf{K}_n+\boldsymbol{\tau}_j\cdot\Delta\mathbf{K}_n+\Delta\boldsymbol{\tau}_j\cdot\mathbf{K}_n+
\boldsymbol{\tau}_j\cdot\mathbf{q}.
$$
The exponent is next expressed to first-order on $\Delta \mathbf{\tau}$:
\begin{eqnarray}
e^{i(\boldsymbol{\tau}_j\cdot\mathbf{K}_n+\boldsymbol{\tau}_j\cdot\Delta\mathbf{K}_n+\Delta\boldsymbol{\tau}_j\cdot\mathbf{K}_n+
\boldsymbol{\tau}_j\cdot\mathbf{q})}\simeq \nonumber\\
ie^{i\boldsymbol{\tau}_j\cdot\mathbf{K}_n}\boldsymbol{\tau}_j\cdot\mathbf{q}+
e^{i\boldsymbol{\tau}_j\cdot\mathbf{K}_n}[1+i(\boldsymbol{\tau}_j\cdot\Delta\mathbf{K}_n+\Delta\boldsymbol{\tau}_j\cdot\mathbf{K}_n)].
\end{eqnarray}
Carrying out explicit calculations one sees that:
\begin{equation}\label{eq:cancellation}
\sum_{j=1}^3e^{i\boldsymbol{\tau}_j\cdot\mathbf{K}_n}[1+i(\boldsymbol{\tau}_j\cdot\Delta\mathbf{K}_n+\Delta\boldsymbol{\tau}_j\cdot\mathbf{K}_n)]=0.
\end{equation}

For example, at $K=\mathbf{K}_2$ we have:
$$
\left[1+\frac{4i\pi(\Delta \tau_{1x}+\Delta \tau_{2x}+\Delta \tau_{3x})}{9a_0}\right](1+e^{\frac{2\pi i}{3}}-e^{\frac{\pi i}{3}}),
$$
with phasors adding up to zero. Similar phasor cancelations occur at every other $K-$point.

The term linear on $\Delta \mathbf{K}_n$ on Eqn.~\ref{eq:cancellation} cancels out the fictitious $K-$point dependent gauge fields proposed in Ref.~\cite{Kitt2012}, which originated from the term linear on $\Delta \mathbf{\tau}_j$ on this same equation. Equation~\eqref{eq:tbh} takes the following form to first order at $\mathbf{K}_2$ in the low-energy regime:
\begin{eqnarray}\label{eq:ps1}
\mathcal{H}_{ps}=&
\left(
\begin{smallmatrix}
0 & t\sum_{j=1}^3ie^{-i\mathbf{K}_2\cdot\boldsymbol{\tau}_j}\boldsymbol{\tau}_j\cdot\mathbf{q}\\
-t\sum_{j=1}^3ie^{i\mathbf{K}_2\cdot\boldsymbol{\tau}_j}\boldsymbol{\tau}_j\cdot\mathbf{q} & 0
\end{smallmatrix}
\right)\nonumber\\
+&\left(
\begin{smallmatrix}
E_{s,A} & -\sum_{j=1}^3\delta t_je^{-i\mathbf{K}_2\cdot\boldsymbol{\tau}_j}\\
-\sum_{j=1}^3\delta t_je^{i\mathbf{K}_2\cdot\boldsymbol{\tau}_j} & E_{s,B}
\end{smallmatrix}
\right),
\end{eqnarray}
 with the first term on the right-hand side reducing to the standard pseudospin Hamiltonian in the absence of strain. The change of the hopping parameter $t$ is related to the variation of length~\cite{vozmediano,Ando2002}:
\begin{equation}
\delta t_j=-\frac{|\beta| t}{a_0^2} \boldsymbol{\tau}_j\cdot\Delta\boldsymbol{\tau}_j.
\end{equation}
This way Eqn.~\eqref{eq:ps1} becomes:
\begin{eqnarray}
\mathcal{H}_{ps}=
\hbar v_F\boldsymbol{\sigma}\cdot \mathbf{q}
+\left(
\begin{smallmatrix}
E_{s,A} & f_1^*\\
f_1 & E_{s,B}
\end{smallmatrix}
\right),
\end{eqnarray}
with $f_1^*=\frac{|\beta|t}{2a_0^2}
[2\boldsymbol{\tau}_3\cdot\Delta\boldsymbol{\tau}_3
-\boldsymbol{\tau}_1\cdot\Delta\boldsymbol{\tau}_1
-\boldsymbol{\tau}_2\cdot\Delta\boldsymbol{\tau}_2
+\sqrt{3}i(\boldsymbol{\tau}_2\cdot\Delta\boldsymbol{\tau}_2-\boldsymbol{\tau}_1\cdot\Delta\boldsymbol{\tau}_1)]$, and $\hbar v_F\equiv
\frac{\sqrt{3}a_0t}{2}$.
The parameter $f_1$ can be expressed in terms of a vector potential: $A_s$ $f_1=-\hbar v_F\frac{eA_s}{\hbar}$. This way:
\begin{eqnarray}\label{eq:Asdiscrete}
\small
A_s&=-\frac{|\beta|\phi_0}{\pi a_0^3}[
\frac{2\boldsymbol{\tau}_3\cdot\Delta\boldsymbol{\tau}_3
-\boldsymbol{\tau}_1\cdot\Delta\boldsymbol{\tau}_1
-\boldsymbol{\tau}_2\cdot\Delta\boldsymbol{\tau}_2}{\sqrt{3}}\nonumber\\
&-i(
\boldsymbol{\tau}_2\cdot\Delta\boldsymbol{\tau}_2
-\boldsymbol{\tau}_1\cdot\Delta\boldsymbol{\tau}_1)],
\end{eqnarray}
and the diagonal entries\cite{usSSC} in Eqn.~\eqref{eq:tbh} are deformation potentials that arise on a neutral system as it is deformed. They indicate the change of the local electronic density as the system is distorted:
\begin{equation}\label{eq:EsA}
E_{s,A}=-\frac{0.3 eV}{0.12}\frac{1}{3}\sum_{j=1}^3\frac{|\boldsymbol{\tau}_j-\Delta\boldsymbol{\tau}_j|-a_0/\sqrt{3}}{a_0/\sqrt{3}},
\end{equation}
and
\begin{equation}\label{eq:EsB}
E_{s,B}=-\frac{0.3 eV}{0.12}\frac{1}{3}\sum_{j=1}^3\frac{|\boldsymbol{\tau}_j-\Delta\boldsymbol{\tau}'_j|-a_0/\sqrt{3}}{a_0/\sqrt{3}}.
\end{equation}
The deformation potential as expressed above has been taken to linear order on the average bond increase following explicit results from {\em ab-initio} calculations \cite{YWSon}.

The deformation potential can be written in terms of the average ($E_{def}$) and the difference ($E_{mass}$)
between $E_{s,A}$ and $E_{s,B}$ (Eqns.~(\ref{eq:EsA}) and (\ref{eq:EsB})) at any given unit cell:
\begin{equation}\label{eq:oldmass}
E_{def}=\frac{1}{2}(E_{s,A}+E_{s,B}), \text{ and } E_{mass}=\frac{1}{2}(E_{s,A}-E_{s,B}).
\end{equation}
Both quantities are of the order of tens of meVs. It is worth noting that many other teams do not include the deformation potential in their models, even though it can lead to significant changes in the electronic spectrum \cite{usSSC,usRapid}.

We also note that, while the determination of $\mathbf{A}_s$ required the preservation of sublattice symmetry, the diagonal terms $E_{s,A}$ and $E_{s,B}$ are not constrained by this requirement because the diagonal entries in Eqns. 22 and 23 remain real regardless of the actual magnitudes of $\Delta \mathbf{\tau}_j$  and of $\Delta \mathbf{\tau}_j'$.

Equations (21-23) are {\em discrete gauge fields}; this is, they take a single value at any given unit cell. These Equations thus provide an original, discrete, viewpoint --in which one retains atomic positions-- to the issue of gauge fields in two-dimensional materials \cite{usPRB,usSSC,usRapid}.

\begin{figure}[tb]
\begin{center}
\includegraphics[width=0.3\textwidth]{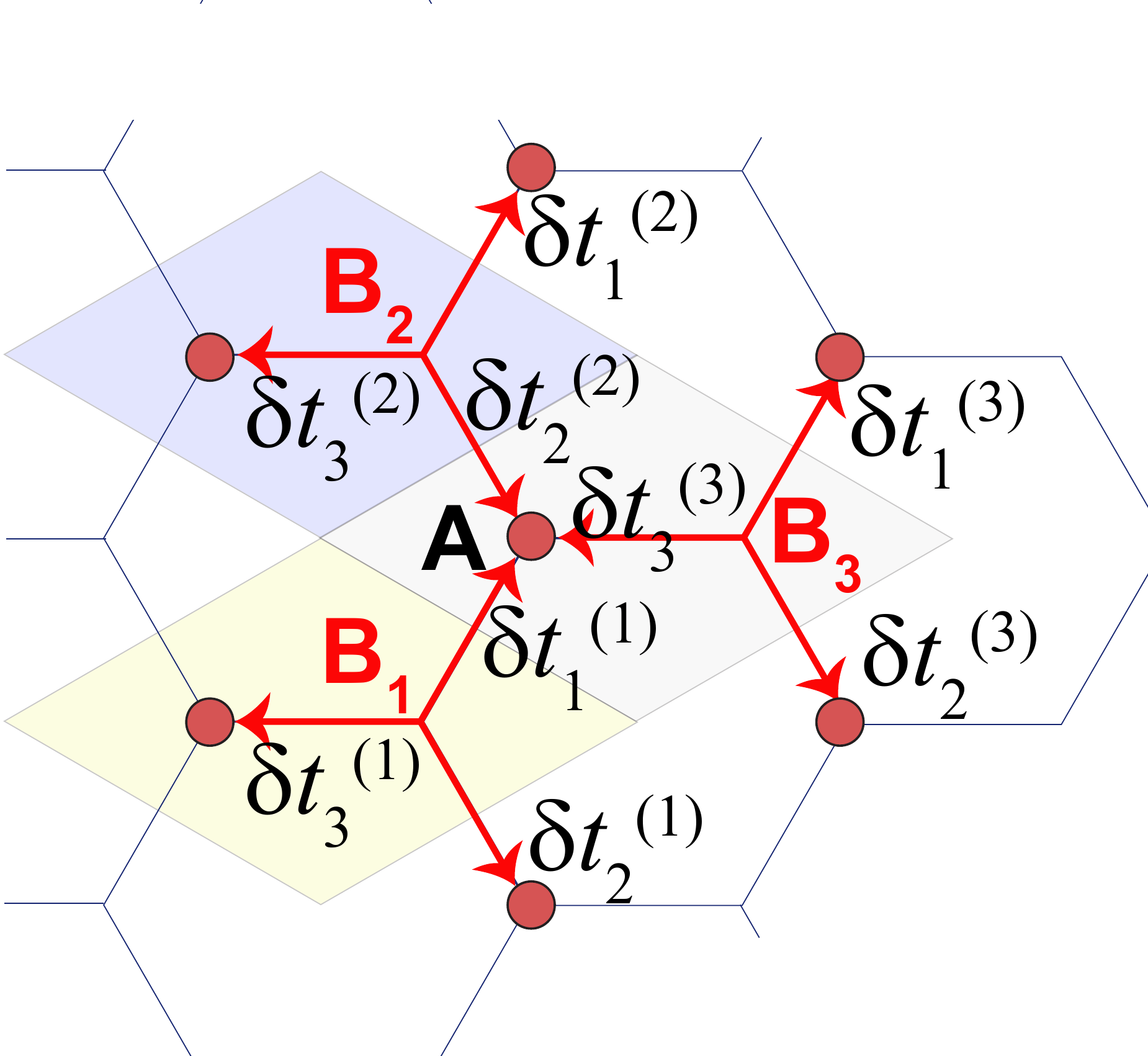}
\end{center}
\caption{(a) The finite-difference curl leading to the pseudomagnetic field $B_s$ [Eq.~\eqref{eq:mass}] is obtained from hoppings among an
atom on the A-sublattice and three neighboring atoms on B-sublattices.}
\label{fig:f4}
\end{figure}

The mass term, Eqn.~\eqref{eq:oldmass} leads to a Zeeman term that arises as a second-order difference relation among potential energies for an atom on the A-sublattice at the $K-$point (c.f, Fig.~\ref{fig:f4}) \cite{usSSC,usRapid,Manes,Sandler}:
\begin{eqnarray}\label{eq:mass}
-\mu_B B_s=&\frac{\sqrt{3}\hbar^2}{m_ea_0^2t}((\delta t_3^{(3)}-\delta t_1^{(3)})-
(\delta t_3^{(2)}-\delta t_1^{(2)})\\+&(\delta t_3^{(3)}-\delta t_2^{(3)})-(\delta t_3^{(1)}-\delta t_2^{(1)})).\nonumber
\end{eqnarray}
Here, $\mu_B$ is the Bohr magneton ($\simeq 5.8 \times 10^{-5}$ eV/Tesla), $\delta t_j^{(n)}$ is the standard change in hopping upon strain
at unit cell $n=1,2,3$ \cite{GuineaNatPhys2010,vozmediano,Ando2002}, and $\frac{\sqrt{3}\hbar^2}{m_ea_0^2t}\simeq 2.5$. The pseudomagnetic field
$B_s$ changes sign at the B-sublattice and/or at the $K'$ point. $E_s$ is the average deformation potential at a
given unit cell~\cite{usSSC} arising from the rearrangement of the electron cloud upon strain \cite{Ando2002}.
 Equation~\eqref{eq:mass} provides a ``microscopic'' vehicle to obtain the local magnitude of the pseudo-magnetic field at any given unit cell directly. We refer readers to Publications \cite{usPRB,usSSC,usRapid} for explicit calculations of gauge fields and electronic spectra in graphene membranes with specific shapes.

\subsection{Relation to the continuum formalism}
The continuum limit is achieved when $\frac{|\Delta\boldsymbol{\tau}_j|}{a_0}\to 0$ (for $j=1,2,3$). We have then (Cauchy-Born rule):
$\boldsymbol{\tau}_j\cdot \Delta \boldsymbol{\tau}_j\to \boldsymbol{\tau}_j\left(
\begin{smallmatrix}
u_{xx}&u_{xy}\\
u_{xy}&u_{yy}
\end{smallmatrix}\right)\boldsymbol{\tau}_j^T$, where $u_{ij}$ are the entries of the strain tensor, and Eqn.~\eqref{eq:Asdiscrete} becomes:
\begin{equation}\label{eq:limit}
A_s\to \frac{|\beta|\phi_0}{2\sqrt{3}\pi a_0}(u_{xx}-u_{yy}-2iu_{xy}),
\end{equation}
as expected \cite{GuineaNatPhys2010,vozmediano}.

$E_{def}$ is an average over changes of distances, and hence reflects the basic form found in terms of the deformation tensor \cite{Ando2002} $E_{def}\propto u_{xx}+u_{yy}$, while $E_{mass}$ takes its continuum form in Eqn.~\eqref{eq:mass}\cite{Manes}; reference \cite{usRapid} contains further details.

We have thus addressed the first question raised in Page 1: We have rewritten the original theory taking into consideration the deformation at individual unit cells.

One of the exciting points of graphene is that it furnishes a field theory for Dirac electrons in 2+1 dimensions. The theory we work with involves fermions having a (pseudo-)spin arising from the $\pi-$electrons on two inequivalent sites on a honeycomb lattice. Unlike an intrinsic spin, a unit cell can be thought as a plaquette that has a finite spatial extent. Each of these plaquettes can be assigned two integer indexes $(i,j)$ that hence furnish a discrete lattice. On a crystal with N atoms, there are N/2 such unit cells that hence define a discrete space. As we work on directly on this discrete space, we say that we realize a discrete field theory. The next point to cover is the discrete geometry in which this lattice gauge theory takes place.

\section{The discrete geometry of two-dimensional materials}
Similar to the statements made in previous section, graphene's geometry is commonly studied in terms of the continuous displacement field $\mathbf{u}\equiv u_{\alpha}(\xi^1,\xi^2)$ as well. Specifically, on thin-plate continuum
elasticity the strain tensor is
$u_{\alpha\beta}=(\partial_{\alpha}u_{\beta}+\partial_{\beta}u_{\alpha}+ \partial_{\alpha}u_{\gamma}\partial_{\beta}u_{\gamma} + \partial_{\alpha}z\partial_{\beta}z)/2$,
with $z$ an out-of-plane elongation \cite{Pereira1,GuineaNatPhys2010,vozmediano,Ando2002,deJuanPRL2012,Manes,deJuanPRB,Dejuan2011,Kitt2013,Peeters3,Peeters4,Naumis1}.
 There, differential geometry and mechanics couple as:
\begin{equation}\label{eq:contgeo}
g_{\alpha\beta}=\delta_{\alpha\beta}+2u_{\alpha\beta},\qquad \text{ }k_{\alpha\beta}=\hat{\mathbf{n}}\cdot \frac{\partial \mathbf{g}_{\alpha}}{\partial \xi^{\beta}},
\end{equation}
where $\mathbf{g}_{\alpha}(\xi^1,\xi^2)$ is a tangent vector field, $\delta_{\alpha\beta}$ is the reference (flat) metric
and $\hat{\mathbf{n}}=\frac{\mathbf{g}_{\xi^1}\times \mathbf{g}_{\xi^2}}{|\mathbf{g}_{\xi^1}\times \mathbf{g}_{\xi^2}|}$ is the local normal \cite{vozmediano,deJuanPRL2012,deJuanPRB,Manes,Dejuan2011}. However, peculiarities of how graphene ripples \cite{ChenNatureNano,Biro,M2,M4,Dumitrica2011}, slides and adheres \cite{Biro,Kitt2} may be beyond first-order continuum elasticity.

We have investigated alternative geometrical frameworks to deal with discrete atomistic surfaces. This is an important endeavor because geometry is behind the spin diffusion in rippled
 graphene \cite{Ando2000,Huertas-Hernando}, behind the chemical properties of conformal (non-planar) two-dimensional crystals \cite{ACSNano}, and may even
 herald the strain engineering of two-dimensional crystals with atomistic defects, an area completely unexplored so far.

 Such discrete geometry exists \cite{ACSNano,usRapid,newest}. There, the Wigner-Seitz/Voronoi unit cells that span a locally-evolving area $A_p$ are the underlying discrete geometrical objects, and the atomistic information is always
 preserved. The discrete formalism for geometry rests on interatomic
 distances without a mediating continuum, just as the theory for the electronic response of Dirac fermions did in previous section.
In what follows, we review the tools for geometrical analysis and study the local geometry of rippled
 graphene \cite{Fasolino1}.

\begin{figure}[tb]
\begin{center}
\includegraphics[width=0.3\textwidth]{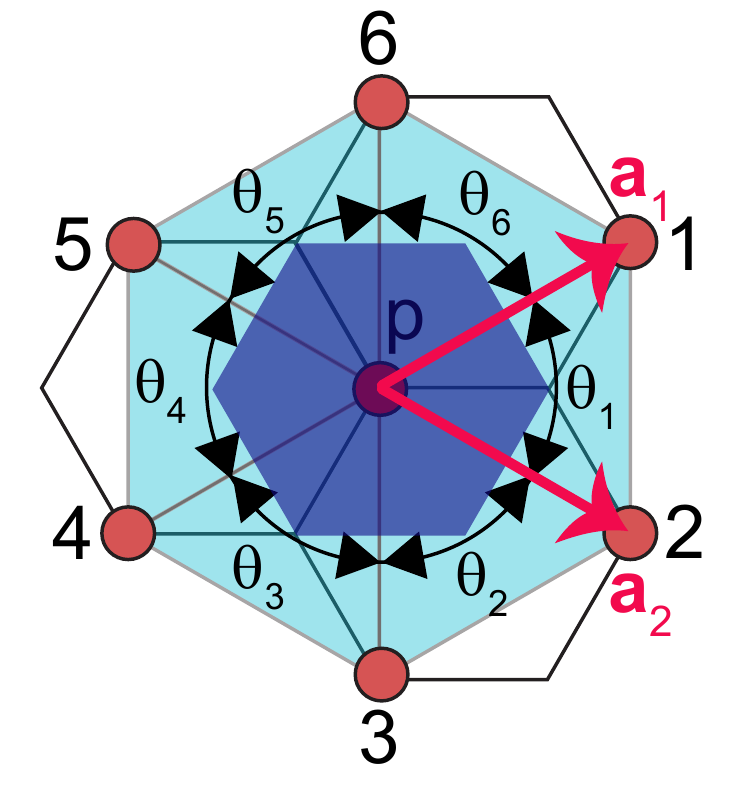}
\end{center}
\caption{Schematics of the parameters employed to determine the local discrete geometry for graphene. $A_p$ is the small hexagon colored in marine blue representing the area of a (Voronoi) unit cell, and the angles $\theta_i$ add up to 2$\pi$ on a flat local surface.}
\label{fig:f3new}
\end{figure}

The four invariants that determine a local shape arise from the metric ($g$) and curvature ($k$) tensors as follows \cite{ACSNano,usRapid}:
\begin{equation}\label{eq:eq1}
\text{Tr}(g), \/\/ \text{Det}(g),\/\/ H\equiv \text{Tr}(k)/2\text{Tr}(g)\text{, }K\equiv\text{Det}(k)/2\text{Det}(g),
\end{equation}
where Tr (Det) stands for the trace (determinant), $H$ is the mean curvature and $K$ is the Gaussian curvature, respectively.

The discrete metric is defined from the local lattice vectors $\mathbf{a}_{\alpha}$ \cite{ACSNano,usRapid}
$g_{\alpha\beta}=\mathbf{a}_{\alpha}\cdot \mathbf{a}_{\beta}$, and the discrete Gauss curvature ($K_D$) originates
from the angle defect \cite{ACSNano,usRapid,TomanekPRB,Math1,Math2,chinese}:
\begin{equation}\label{eq:DGB}
K_D=(2\pi-\sum_{i=1}^6\theta_i)/A_p.
\end{equation}
Here $\theta_i$ ($i=1,...,6$) are angles between vertices and $A_p$ will be defined below; see Fig.~\ref{fig:f3new}.
The {\em Voronoi tessellation} generalizes the Wigner-Seitz unit cell to conformal two-dimensional geometries \cite{ACSNano,usRapid}.

 The discrete mean curvature $H_D$ measures relative orientations of edges and normal vectors along a closed path:
\begin{equation}\label{eq:Hdiscrete}
H_{D}=\sum_{i=1}^6 \mathbf{e}_i\times(\boldsymbol{\nu}_{i,i+1}-\boldsymbol{\nu}_{i-1,i})\cdot \hat{\mathbf{n}}/4A_p.
\end{equation}
Here, $\mathbf{v}_i$ is the position of atom $i$ on sublattice $A$, and $\mathbf{e}_i=\mathbf{v}_i-\mathbf{v}_p$ is the {\em edge} between points
$p$ and $i$ (note that $\mathbf{a}_{1(2)}=\mathbf{e}_{1(2)}$). $\boldsymbol{\nu}_{i,i+1}$ is the normal to edges $\mathbf{e}_i$ and $\mathbf{e}_{i+1}$ ($i$
is a cyclic index), and  $\hat{\mathbf{n}}=\frac{\sum_{i=1}^6\boldsymbol{\nu}_{i,i+1}A_i}{\sum_{i=1}^6A_i}$ is the area-weighted normal
with $A_i=|\mathbf{e}_i\times \mathbf{e}_{i+1}|/2$ \cite{Math1}, and $A_p=\frac{1}{3}\sum_{j=1}^6A_j$.

  The discrete metric and curvatures furnish geometry consistent with a crystalline structure and they lead to the faithful characterization of graphene's
  morphology beyond the effective-continuum paradigm, Eqn.~\eqref{eq:contgeo}. This is advantageous when the atomic conformation is known from molecular dynamics   (e.g, \cite{Fasolino1}) or experiment (e.g., \cite{MolecularGraphene}) because: (i) fitting of the atomic lattice to an effective continuum is not needed
  any more, (ii) the Chemistry of conformal graphene can be addressed from the discrete geometry \cite{ACSNano} and, since atoms are always
  available, (iii) the discrete theory brings new insights and understanding into the physical theory (e.g., non-preservation of sublattice symmetry, the
  form of gauge fields \cite{usPRB}, the creation of mass from strain \cite{usSSC,Manes}). We emphasize that the discrete geometry is accurate regardless of elastic regime, hence
  it can be used to verify whether the conditions for continuum elasticity hold in the problem at hand.

\subsection{The geometry of rippled graphene}
The importance of a sound geometrical framework is motivated by rippled graphene. We contrast ripples created
by thermal fluctuations \cite{Fasolino1} with those created at low temperature due to edges. These two mechanisms lead to different types
of geometries (hence different magnitudes of strain-derived gauges). In a system with periodic boundary conditions, thermal fluctuations
create significant changes in interatomic distances (i.e., in the metric) \cite{Fasolino1}  and --as the boundaries are fixed-- such increases on interatomic
distances produce out-of-plane deformations (i.e., rippling).

Now consider a square graphene sample with about three million atoms, in which strain was relieved at the low temperature of 1 Kelvin.
The resulting membrane is shown in Fig.~5(a), where colors indicate varying heights across the sample \cite{usSSC}. Ripples in Ref.~\cite{Fasolino1} originate
from {\em increases} in the metric. On the other hand, the white margin in
between the ``rippled'' (curved) sample and the (yellow) exterior frame highlights an apparent {\em contraction} of our finite sample
when seen from above.

\begin{figure}[tb]
\begin{center}
\includegraphics[width=0.47\textwidth]{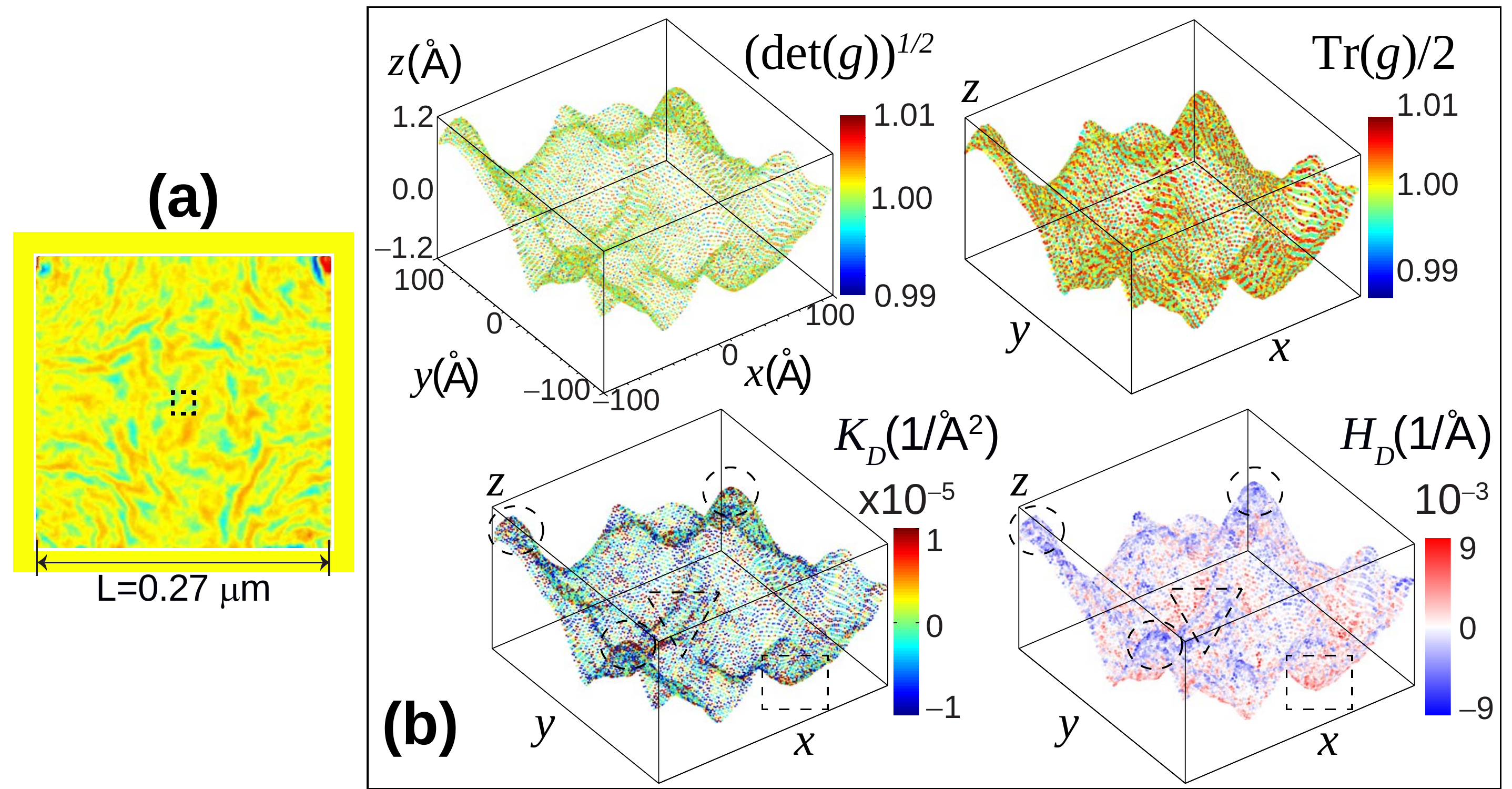}
\end{center}
\caption{(Color online) (a) Creation of ripples by cutting a square with side $L=0.27 \mu m$ at 1 Kelvin: The membrane
trades a planar configuration for a rippled one. (b) Geometrical invariants within the dashed square shown in (a). 
}\label{fig:f2}
\end{figure}

The details of this geometry are shown in Fig. 5(b): Det($g$) and Tr$(g)$ are unity
almost everywhere (yet there are significant random fluctuations driving the scales). The metric tells us that the membrane
does not contract, and its area thus remains almost unchanged. We show in Fig.~5(b) the discrete
curvatures, highlighting cusps by ovals, valleys by squares, and ridges by triangles.
Cusps and valleys have the largest Gaussian curvature $K_D$ (deep red), while
ridges have the smallest one (deep blue). As expected, the mean curvature $H_D$ takes its largest (smallest) value at valleys (cusps) and alternates sign around
ridges. The curvature --without metric increases-- explains the white margins on Fig.~5(a).

 The discrete geometry reflects the mechanism leading to ripple formation, thus highlighting the virtue
 of a geometry that originates from atoms. An accurate determination of $H_D$ is important since $H_D$ leads to spin diffusion in rippled
 graphene \cite{Ando2000,Huertas-Hernando}. Though much has been said about ripples, no geometrical study with the accuracy provided in Ref.~\cite{ACSNano,usRapid} exists.

 The starting point in the continuum theory is a flat metric $\delta_{\alpha\beta}$. There, a non-zero curvature directly leads to
 increases in interatomic distances Eqn.~\eqref{eq:contgeo}, and a non-zero height is directly identified with a non-zero strain-derived gauge. A question
 then arises whether the sample under study actually obeys Eqn.~\eqref{eq:contgeo}. The situation shown in Fig.~5 is a counterexample to the geometry inferred
 from Eq.~\eqref{eq:contgeo}, because the metric is almost constant, even though the height profile $z$ is clearly non-flat. Gauge fields
 for similar samples were reported in Ref.~\cite{usSSC}. Fig.~5 represents the accurate geometrical characterization of rippled graphene down to the atomic level. We studied the geometry of graphene under load, providing lattice gauge fields and electronic spectra, in Ref.~\cite{usRapid}.

We next address another aspect of conformal two-dimensional materials: The potential increase in chemical reactivity under non-planar, conformal shapes.

\subsection{Chemical measures and geometry}

 We first provide two known chemical measures for carbon-based materials:

\begin{figure*}
\begin{center}
\includegraphics[width=0.75\textwidth]{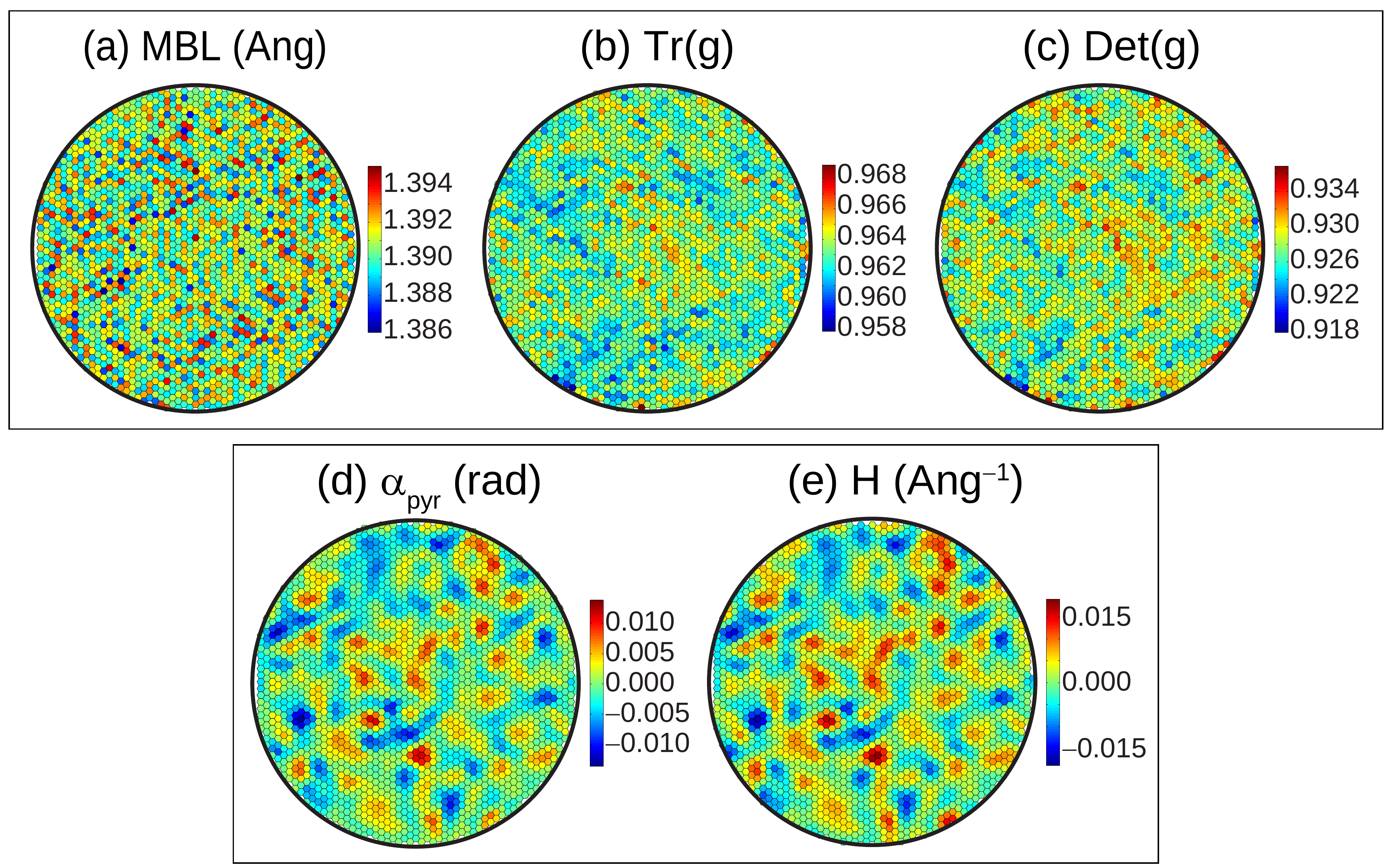}
\end{center}
\caption{Chemical measures and the local geometry of a rippled two-dimensional crystal. While no direct correlation can be drawn among MBL in (a) and metric measures in (b-c), the piramidalization $\alpha_{pyr}$ in (d) is directly proportional to the mean curvature $H$ in (e). Actual atomic bonds are seen in black \cite{ACSNano}.}
\end{figure*}

\begin{enumerate}
\item{}Mean Bond Length (MBL): Aromaticity is not a directly measurable property and hence it cannot be defined unambiguously. Yet, the structural representation accommodating the maximum number of Clar sextets best represents chemical and physical properties, and Clar sextet migration increases chemical reactivity. How aromatic is conformal graphene \cite{Fasolino1}? Can rippling be explained in terms of the creation of ``aromatic domains''? The mean bond length (MBL) is defined as follows \cite{ACSNano,Wu,r49}:
\begin{equation}\label{eq:MBL}
MBL=\bar{a}_{CC}=\frac{1}{6}\sum_{i=1}^6 {a}_{CC,i},
\end{equation}
where ${a}_{CC,i}$ are bond lengths on a closed loop \cite{ACSNano}. According to Ref.~\cite{r49}, $MBL$ is a reliable tool for analysis of large aromatic systems.
\item{}The angle $\theta_{\sigma\pi}$ between the bonds and the normal vector $\hat{\mathbf{n}}$ at atom $p$ has a single value $\theta_{\sigma\pi}$ under a spherical geometry \cite{Haddon} and $\theta_{\sigma\pi}=\pi/2$ on a flat surface. $\theta_{\sigma\pi}$ can be generalized for arbitrary geometries as an average:
\begin{equation}\label{eq:pa}
\bar{\theta}_{\sigma\pi}\equiv\frac{1}{3}\sum_{i=1}^3\theta(i)_{\sigma\pi},
\end{equation}
with $\theta(i)_{\sigma\pi}$ the angle among a bond vector and the local normal. Equation \eqref{eq:pa} takes its usual form for fullerenes, where $\theta(i)_{\sigma\pi}=\theta_{\sigma\pi}$ for all bonds \cite{Haddon}. The pyramidalization angle $\alpha_{pyr}$ was defined by Haddon  as follows \cite{Haddon}:
\begin{equation}\label{eq:pyr}
\alpha_{pyr}=\bar{\theta}_{\sigma\pi}-\pi/2.
\end{equation}
\end{enumerate}

The degree of sensitivity of MBL with respect to fluctuations on interatomic distances makes a direct correlation difficult \cite{Fasolino1,usRapid,usSSC}. In Figures 5(a-c) we contrast MBL with  $Tr(g)$ and $det(g)$ (we will not display the determinant of the metric tensor in further figures). Details of the creation of the rippled structure can be found in prior work \cite{usPRB,usSSC,usRapid}. Recalling the notion that pristine graphene has equal bond lengths, Fig.~6(a) indicates that atomistic fluctuations will have a bearing on the aromatic behavior of rippled samples; this concept has not been discussed before, nor its ramifications.

 On the other hand, there exists a remarkably simple, one-to-one correlation among the pyramidalization angle and the mean curvature (sign included) for all the systems studied, as already evident from Fig.~6(d-e):
\begin{equation}\label{eq:correlation}
\alpha_{pyr}\text{ (in rads) }\simeq 1\times H_D \text{ (in \AA$^{-1}$).}
\end{equation}
Equation \eqref{eq:correlation} is an interesting result because it relates a commonly used angular measure for orbital hybridization and chemical reactivity with the mean curvature. This result was hidden in plain sight; this shows once again how the discrete geometry makes plenty of sense.

$\alpha_{pyr}$ is a signed quantity, as follows: Direct inspection of Eqn.~\eqref{eq:pyr} indicates that $\alpha_{pyr}$ will be positive for a bulge, and negative for a sag. Similarly, the mean curvature $H_D$ --Eqn.~\eqref{eq:Hdiscrete}-- is a vector quantity projected onto the local normal; the relative orientation of the normal (facing ``up'' or ``down'') confers $H_D$ with a sign as well. (Geometrically speaking, one sees that radius of curvature changes sign for a bulg or a sag, so $H_D$ must be signed.) But the correspondence goes beyond the sign. The cross products on $H$ --Eqn.~\eqref{eq:Hdiscrete}-- confers an additional sinusoidal function, which approximates as the angle rather well up to 20 degrees (0.35 rad), within the range of all pyramidalization angles we saw. The correlation given by Eqn.~\eqref{eq:correlation} is remarkable as it informs our intuition concerning hybridization, thus making the mean curvature a {\em direct tool} for analysis of hybridization and chemical reactivity for two-dimensional systems with $s$ and $p$ electrons.

Additional results in Ref.~\cite{ACSNano} include the discussion of a geometry of systems with topological defects such as fullerenes \cite{Kroto,fullerenes1,fullerenes2,fullerenes3,fullerenes4,fullerenes5,fullerenes6}, Schwarzites \cite{H1}, ionic crystals \cite{WalesPRB2009,WalesPRL2013}, and other hexagonal systems with atomistic defects.

This concludes the main discussion on this contribution. As advances in two-dimensional materials are occurring at great speed, we end this work briefly highlighting some contributions to such endeavor arising from our group.

\section{Brief account of additional contributions on graphene and on other two-dimensional materials}

The tools employed in previous sections are a combination of electronic structure within a tight-binding approach, molecular dynamics, and a discrete geometry.

But the group has been invested on models of electron transport that capture the electronic structure of graphene and metal leads with a Green's function approach that is coupled to an {\em ab-initio} electron Hamiltonian \cite{t1,t2,t3}; these models provide noise features that reproduce experimental features \cite{Marcus} not seen in more basic theoretical models \cite{Beenakker}.

Another result from our group that is becoming relevant and is worth mentioning concerns the experimental observation on an STM of a bulk material through graphene and its theoretical confirmation \cite{NL2012}; the result is interesting because it can be used, for instance, to study the surface of black phosphorus through graphene or through hexagonal boron nitride monolayers.\cite{PRB2015}

 We briefly discuss in what follows our contributions to other two-dimensional materials. The techniques employed in these studies combine {\em ab-initio} methods, tight-binding models, and the discrete geometry.

\subsection{Stanene}
\begin{figure}[tb]
\includegraphics[width=0.475\textwidth]{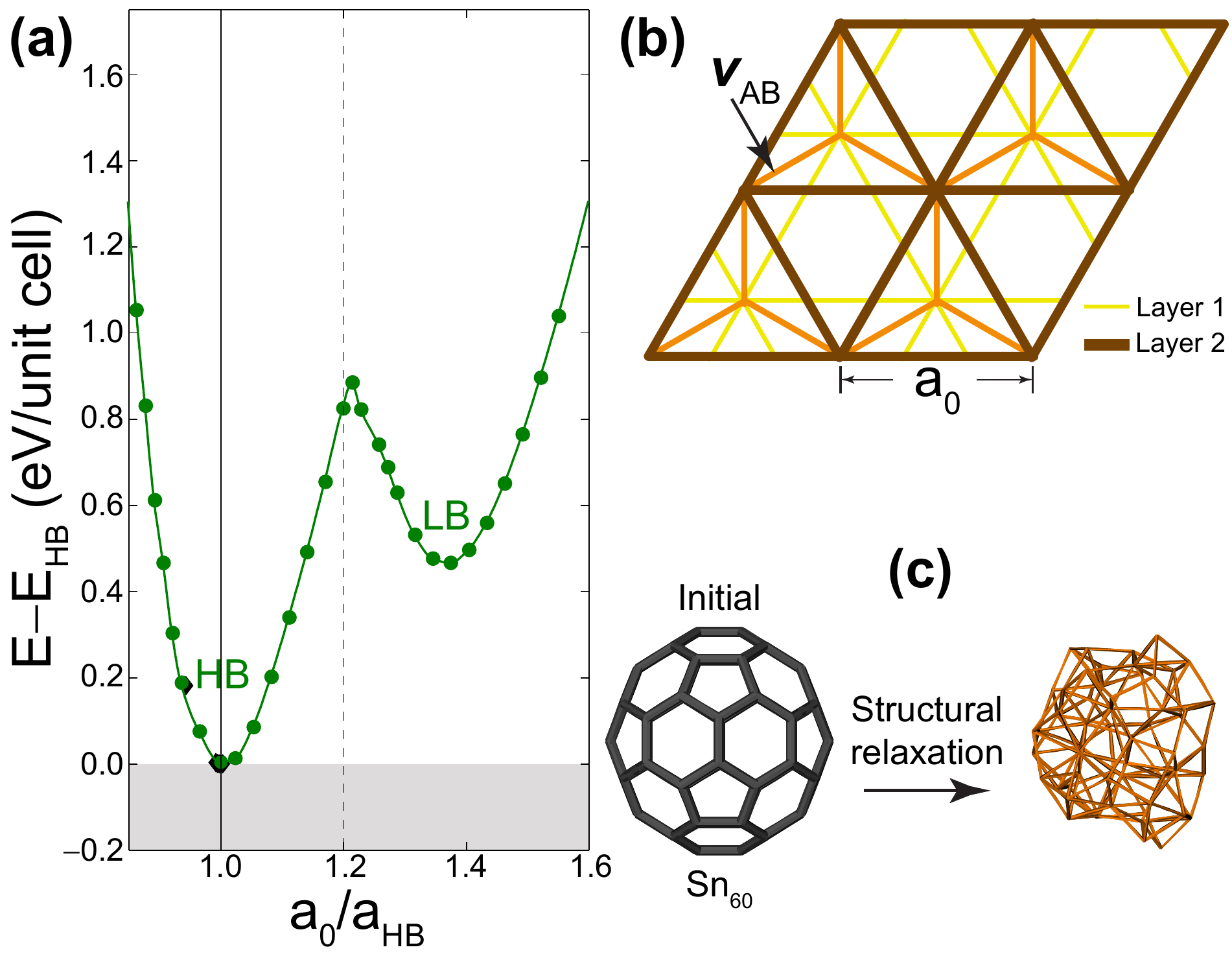}
\caption{(a) The high-buckled (HB) phase is more stable than the low-buckled (LB) phase, making freestanding stanene ~\cite{jpc2,stanene} metastable. (b) The actually stable, high-buckled phase, is a HCP bilayer with trivial electronic properties (it is a metal). (c) As seen on this one-minute-long structural optimization, stanene does not realize topological fullerenes \cite{circular} either.}\label{fig:F1p}
\end{figure}

Proceeding by direct analogy to silicene and germanene~\cite{Ciraci2009}, known studies of the electronic properties of two-dimensional
tin \cite{jpc1,jpc2,stanene} have been performed under the implicit assumption  that the HB phase is not viable. Contrary
to this assumption, we determined using {\em ab-initio} methods that the HB two-dimensional structures of heavy column-IV elements tin and lead
are stable and lower in energy than their LB counterparts (c.f., Fig.~\ref{fig:F1p}a), thus representing the true optimal structures of these two-dimensional systems \cite{Pablo2}. The HB structure is a hexagonal close-packed bilayer (c.f., Fig.~\ref{fig:F1p}a).

 Haldane's honeycomb model has been studied in closed geometries \cite{circular} and one of the many candidates for its practical realization is
 LB tin (stanene). Unfortunately, a fullerene-like Sn$_{60}$ is not stable (Fig.~\ref{fig:F1p}c) so tin and lead are no-go elements for topological fullerenes \cite{Pablo2}.

We also determined \cite{Pablo2} that the optimal phase of two-dimensional fluorinated stanene is not analogous to tetrahedrally-coordinated
 graphane \cite{graphane} as it was postulated in Refs.~\cite{jpc2,stanene} either. There is no indication for
 tetrahedral coordination of tin atoms in bulk fluorinated tin ~\cite{mcdonald} and tetrahedral coordination~\cite{jpc2,stanene} does not yield the
 most stable two-dimensional fluorinated tin either.

\begin{figure}[tb]
\includegraphics[width=.475\textwidth]{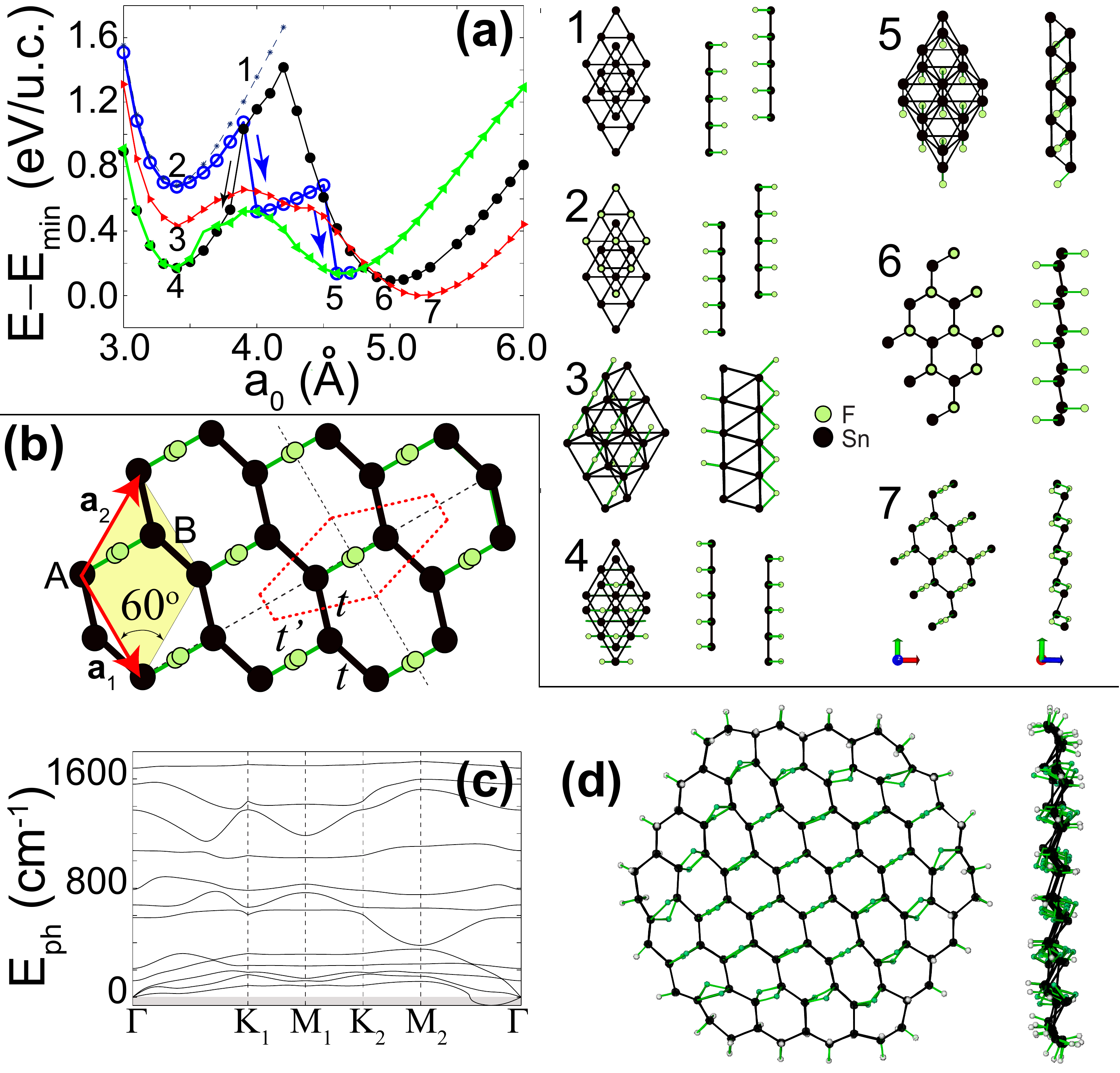}
\caption{(a) Phases of two-dimensional fluorinated tin; structures shown to the right. (b) Symmetries of the most stable structure (7),
depicting
triangular (dashed) and Wigner-Seitz (within dotted perimeter) unit cells, the two symmetry axes, and the two Sn sublattices $A$ and $B$.
Structural
stability is demonstrated by (c) phonon dispersion curves and (d) the structural stabilization of a finite-size sample.}\label{fig:F4p}
\end{figure}

Indeed, as seen in Fig.~\ref{fig:F4p}a, the phase space for decorated two-dimensional tin is larger than originally anticipated: The
graphane-like  phase \cite{jpc2,stanene} realizes the metastable minima labeled {\bf 6} that turns into phase {\bf 4} upon in-plane compression.
  In the optimal structure, {\bf 7}, four-fold coordinated Sn atoms form a sequence of parallel zig-zag one-dimensional chains
  with two fluorine atoms mediating interactions among neighboring Sn chains.  The structure is realized on a triangular lattice with $a_0=5.230$ \AA{}
  [Fig.~\ref{fig:F4p}b]. The Wigner-Seitz unit cell is within the dotted area in Fig.~\ref{fig:F4p}b, where the symmetry axes are shown as well.

\begin{figure}[tb]
\includegraphics[width=.475\textwidth]{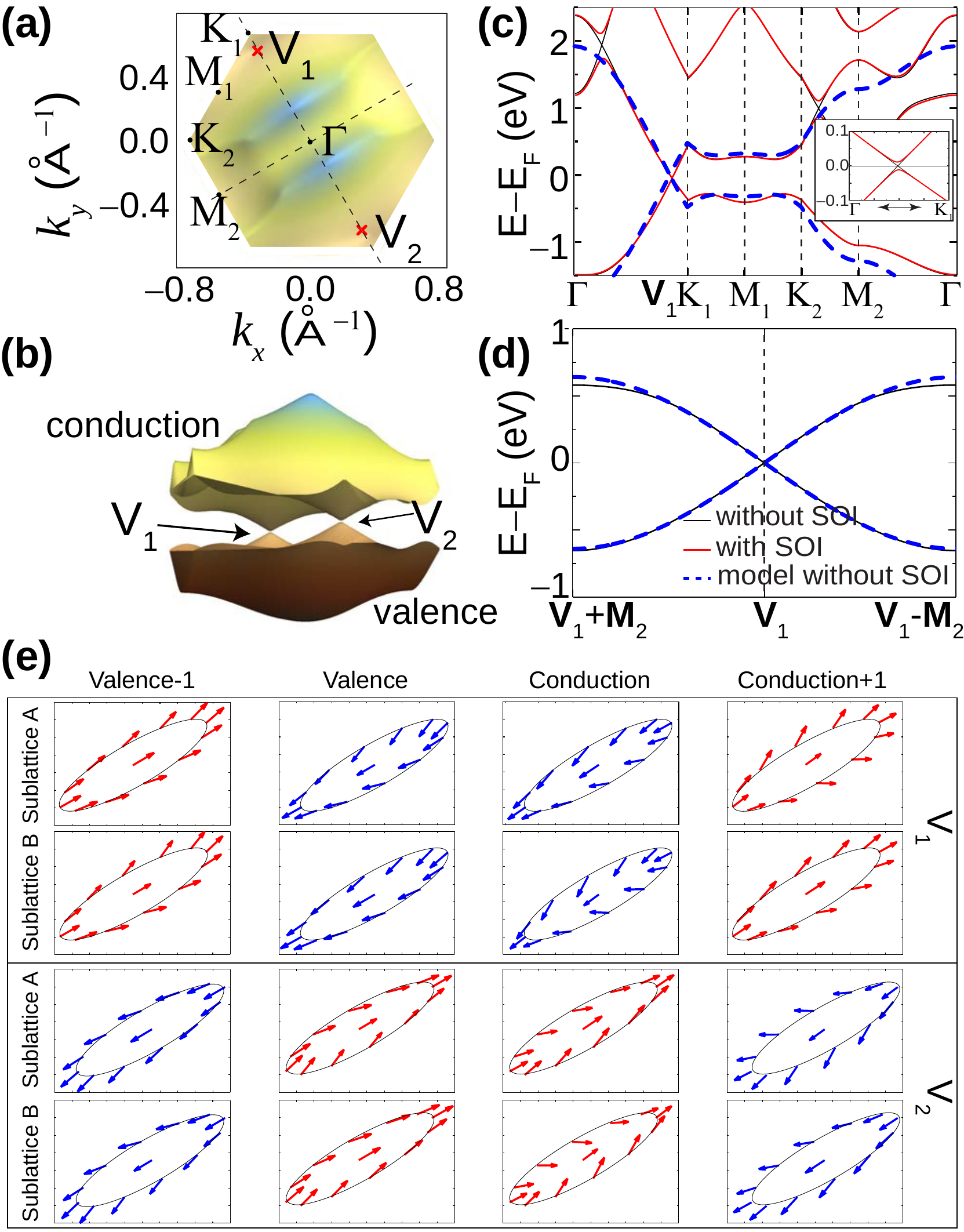}
\caption{(a) Conduction band on the first Brillouin zone, highlighting high-symmetry points and locations of valleys {\bf V}$_1$ and
{\bf V}$_2$ away from the K- and K'-points. (b) The two valleys on the Brillouin zone arise from the two-fold symmetry of the atomic structure. (c-d)
Band structures along high-symmetry lines, including a two-band tight-binding fit. (e) Spin texture resolved over valley ($\boldsymbol{\tau}$), energy,
and sublattice ($\boldsymbol{\sigma}$) degrees of freedom. (The spin projection onto the $z-$axis is of the order of 1\% at most.)}\label{fig:F5p}
\end{figure}

 The first Brillouin zone in Fig.~\ref{fig:F5p}(a) shows a top view of the conduction band and the high-symmetry points in momentum space. As seen in Fig.~\ref{fig:F5p}(b), the
 arrangement of parallel 1D Sn wires gives rise to an electronic structure with only two anisotropic Dirac cones on the First Brillouin zone located away
 from the K-points at positions {\bf V}$_1$ and {\bf V}$_2$ $=\pm$0.85{\bf K}$_{1}$, respectively. From now on we identify the $x-$axis with the line
 joining tin atoms across fluorine bridges. The Fermi velocity is close in magnitude to that of graphene and it is anisotropic: $v_{Fy}=5.4\times 10^{5}$
 m/s [Fig.~9(c)], and $v_{Fx}=2.1\times 10^{5}$ m/s [Fig.~\ref{fig:F5p}(d)] and a $2\Delta= 0.02$ eV gap opens due to SOI,\; five times larger
 than the intrinsic gap due to SOI in graphene \cite{Huertas-Hernando}. Phase {\bf 6} transitions from a topological insulator to a trivial
 insulator \cite{stanene}, but the electronic structure of the optimal phase remains robust under larger isotropic strain.

The electronic dispersion in Fig.~\ref{fig:F5p}(b-d) can be understood in terms of a $2\times 2$ $\pi-$electron tight-binding Hamiltonian \cite{Gilles} in
 which an effective coupling $t'$ is set among the tin atoms originally linked by fluorine bridges [thin bonds on Fig.~\ref{fig:F4p}(b)],
 and $t$ is the coupling among actual Sn-Sn atoms [thick bonds on Fig.~\ref{fig:F4p}(b)]. Using interatomic distances among Sn atoms from Table I we obtain
 the blue dashed lines in Fig.~\ref{fig:F5p}(c,d) with $t=0.8$ eV and $t'=\frac{v_{Fx}}{v_{Fy}}t$ which reproduce first-principles results.

To account for SOI, we realize an oblate low-energy Dirac-Hamiltonian at the vicinity of the {\bf V}$_{1,2}$ points. The numerical results on Fig.~\ref{fig:F5p}(e) are consistent with a coupling $\tau_z\sigma_x s_x$ \cite{Pablo2}. Indeed, eigenvectors of $\tau_z\sigma_x s_x$ project spins onto the $-x$, $+x$, $+x$, $-x$ axis parallel to the Sn-F bonds, inverting signs at each valley and lacking sublattice polarization, consistently with {\em ab-initio} data [Fig.~9(e)]. Thus, the low-energy dynamics is given by:
\begin{equation*}
 H=-i\hbar\Psi^{\dagger}(v_{Fx}\tau_z\sigma_x\partial_x + v_{Fy}\sigma_y\partial_y)\Psi + \Psi^{\dagger}(\Delta \tau_z\sigma_x s_x)\Psi.
\end{equation*}
An unprecedented specific coupling of momentum --including direction-- with spin oriented along $\hat{\mathbf{x}}$
and valley degrees of freedom is thus realized by the second term in previous equation. The valley degree of freedom can be addressed
by a bias along the $\mathbf{V}_1-\mathbf{V}_2$ axis that breaks inversion symmetry. Similarly,
a magnetic field along the $\hat{\mathbf{x}}$ axis will break time-reversal symmetry, locking the valley and crystal momentum direction at
the $\mathbf{V}_1$, $\mathbf{V}_2$ points. The dynamics invites the use of two-dimensional fluorinated tin for valleytronic applications.

 The structural stability of HB tin and HB lead clearly have fundamental consequences for the practical realization of substrate-free non-trivial
 topological phases based from these elements.

\subsection{Phosphorene}
Studies of planar phosphorene with defects have begun to appear \cite{Jakobson2,Tiling} and single-digit-percent strain on planar black phosphorene induces a ten-fold change on its electronic gap \cite{Peeters}. Curvature can induce strain \cite{Haddon}, and we indicated how to induce a positive Gaussian curvature on black phosphorene, how to characterize such geometry, and how this shape influences the electronic gap. Additionally, we demonstrate that this reduction in the fundamental gap applies to other phosphorene allotropes.

Finite-size phosphorene cones --with a positive Gaussian curvature-- were built by {\em removing} the angular segments seen in Fig.~10, joining atoms along the red lines, and a subsequent atomistic relaxation with molecular dynamics at the {\em ab-initio} level. (An area segment must be {\em added} instead to build structures with negative Gaussian curvature --e.g.; \cite{ACSNano,TomanekPRB}.) The angular sections removed subtend a 46$^\circ$ angle for black phosphorene. To get a suitable joining line for black phosphorene, the starting point is a planar structure with a dislocation at an angle of 26.8$^\circ$ that does not create localized electronic states \cite{Jakobson2} but confers additional structural rigidity. The edges were passivated with Hydrogen atoms, and the conical structure on Fig.~10 contains about five hundred atoms. The electronic gap for the planar flake takes a constant value  of 1.1 eV, as highlighted by the yellow color in the last column of Fig.~10. The magnitude of the gap is about twice its nominal value (0.4 eV in standard density-functional theory \cite{Phosphorene1,TomanekBlue}) due to finite-size effects.

\begin{figure}[h]
\centerline{\includegraphics[width=0.5\textwidth]{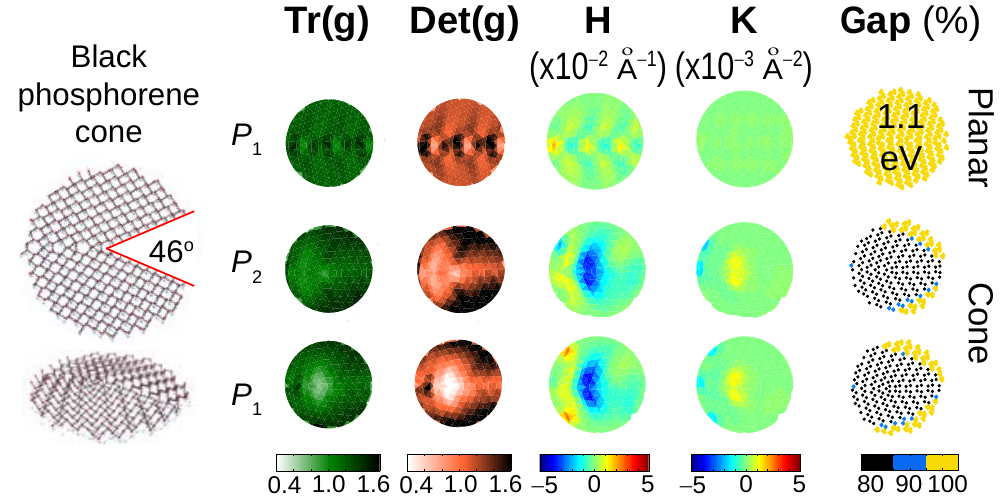}}
\caption{The discrete geometry and the semiconducting gap of black phosphorene cones. The cones are built by removing the indicated segments on an initially planar structure. A remarkable $\sim$20 percent reduction of the semiconducting gap occurs due to the strain induced by the conical shape.}\label{fig:F7}
\end{figure}

 The black phosphorene cone subtends a solid angle of 0.83$\times 2\pi$ radians, thus inducing a positive Gaussian curvature, and it shows compressive strain at the apex as indicated by the white tones in their metric around this point. Black phosphorene has a large structural rigidity due to its ridged structure and as a result the metric and the curvatures in conical structures lack a perfect radial symmetry. The discrete metric and $H_D$ tell us the locations of the dislocation line in the planar structure.

The last column in Fig.~10 depicts the semiconducting gap at each atomic position, and the darkest color indicates a 20\% reduction of the gap with respect to its value in the planar structure, due to the curvature-induced structural compression \cite{Haddon} discussed in previous paragraph. Thus, topological defects can help in tuning the local gap of phosphorene \cite{newest}.

\section{Conclusion}
 In conclusion, we have presented here a number of contributions to two-dimensional materials in which their discrete geometry plays a preponderant role.

\section{Acknowledgments}
We acknowledge the Arkansas Biosciences Institute for financial support. Computations have been carried out at Arkansas and TACC (XSEDE TG-PHY090002).

\end{document}